\documentclass[conference]{IEEEtran}

\usepackage{graphicx}

\ifCLASSINFOpdf
\else
\fi
\begin{document}
\title{A Precision Diagnostic Framework of Renal Cell Carcinoma on Whole-Slide Images using Deep Learning}

% \author{\IEEEauthorblockN{Jialun Wu^{1},
% Haichuan Zhang^{1,2}, 
% Tieliang Gong^{1},
% Chunbao Wang^{3},
% Chen Li^{1,*}}

\author{
\IEEEauthorblockN{Jialun Wu$^{1}$, Haichuan Zhang$^{1,2}$,Zeyu Gao$^{1}$, Xinrui Bao$^{3}$, Tieliang Gong$^{1}$, Chunbao Wang$^{4}$, and Chen Li$^{1,*}$}
\IEEEauthorblockA{$^{1}$School of Computer Science and Technology, Xi'an Jiaotong University, Xi'an, China\\
National Engineering Lab for Big Data Analytics, Xi'an Jiaotong University, Xi'an, China} 
\IEEEauthorblockA{$^{2}$School of Electrical Engineering and Computer Science, Pennsylvania State University, University Park, PA, USA}
\IEEEauthorblockA{$^{3}$School of Automation Science and Engineering, Xi'an Jiaotong University, Xi'an, China}
\IEEEauthorblockA{$^{4}$Department of Pathology, the First Affiliated Hospital of Xi'an Jiaotong University, Xi'an, China\\
Email: andylun96@stu.xjtu.edu.cn}
}


%Telephone: (800) 555--1212, Fax: (888) 555--1212}


\maketitle

\begin{abstract}
Diagnostic pathology, which is the basis and gold standard of cancer diagnosis, provides essential information on the prognosis of the disease and vital evidence for clinical treatment. Tumor region detection, subtype and grade classification are the fundamental diagnostic indicators for renal cell carcinoma (RCC) in whole-slide images (WSIs). However, pathological diagnosis is subjective, differences in observation and diagnosis between pathologists is common in hospitals with inadequate diagnostic capacity. The main challenge for developing deep learning based RCC diagnostic system is the lack of large-scale datasets with precise annotations. In this work, we proposed a deep learning-based framework for analyzing histopathological images of patients with renal cell carcinoma, which has the potential to achieve pathologist-level accuracy in diagnosis. A deep convolutional neural network (InceptionV3) was trained on the high-quality annotated dataset of The Cancer Genome Atlas (TCGA) whole-slide histopathological image for accurate tumor area detection, classification of RCC subtypes, and ISUP grades classification of clear cell carcinoma subtypes. These results suggest that our framework can help pathologists in the detection of cancer region and classification of subtypes and grades, which could be applied to any cancer type, providing auxiliary diagnosis and promoting clinical consensus.
\end{abstract}

% \begin{IEEEkeywords}
% multi-source heterogeneous data, pathology report, whole-slide image.
% \end{IEEEkeywords}


% \fi
\IEEEpeerreviewmaketitle

\section{Introduction}
Renal cell carcinoma (RCC) is one of the most common types of renal cancer, accounting for 80\% of early renal cancer. Renal cancer is the third most common urological tumor after prostate cancer, and bladder cancer \cite{1}. With the development of medical imaging, the detection rate of early kidney cancer is gradually increasing. The most common histopathological type of renal cell carcinoma is clear cell carcinoma (ccRCC), papillary renal cell carcinoma (pRCC), and chromophobe cell carcinoma (chRCC) \cite{2}. In the pathological diagnosis, subtypes and grades of renal cell carcinoma are the critical diagnostic results, different subtypes of renal cell carcinoma can be treated with different regimens (including chemotherapy and targeted therapy), the higher the grading, the worse the prognosis, the greater the possibility of recurrence and tumor metastasis, and the faster the course of the disease.

Pathology is the foundation and gold standard in the medical field. The pathological diagnosis of tumor tissue by pathologists is the basis of treatments and the cornerstone of clinical and drug research. Most carcinoma identification requires microscopy-level image assessment for early tumor discovery and for developing therapies based on diagnostic pathology \cite{3}. Histopathological images have long been central to cancer diagnosis, staging, and prognosis, and pathologists widely use them in clinical practice. Diagnosing pathology slides is a complex task that requires years of pathologist training, digital pathology slides are obtained at very high resolution. However, with the rapid growth of cancer patients, the technical requirements for pathologists also increase. At present, pathological diagnosis is confronted with such problems as uneven distribution of medical resources, huge workload of pathologists, uneven level and serious shortage of doctors engaged in pathological work \cite{4}. The diagnosis and treatment of some diseases often lack precise pathological examination results and are only based on speculation with incomplete evidence. It takes a significant amount of time to train a pathologist and even longer for pathologist to become experienced and be able to diagnose different types of tumor.

As a new tool in the field of pathology, artificial intelligence uses intelligent pathology-assisted diagnosis technology to collect, manage and analyze pathological information, which can help pathologists reduce a lot of workload, effectively improve the efficiency and accuracy of pathological diagnosis,provide better patient treatment and good support in clinical teaching of pathology. The predictive models used by traditional medical image-assisted diagnostic systems rely on the features extracted manually by the pathologists, but the performance is often inadequate for clinical practice. Deep learning (DL) is a powerful method for tumor region detection, subtypes and grades classification of the whole-slide images in digital pathology \cite{5}. It has demonstrated high accuracy and universality in diagnosing the most common human cancers, such as kidney cancer \cite{6}, lung cancer \cite{7}, prostate cancer \cite{8}, colon cancer \cite{9}, breast cancer \cite{10}. Deep learning with convolutional neural networks (CNNs) has been shown to be a powerful algorithm for prompting biomedical image analysis. There have been works using convolutional neural networks \cite{11} for tumor region detection and cell segmentation \cite{12,13}, with gridding the whole-slide images into small patches for processing. A major obstacle to applying deep learning-based algorithms in the analysis of pathological images is the lack of well-annotated training datasets, which require professional pathologists to make a large number of annotations at both patch level and pixel level.

To tackle these problems, we curated a set of pathology images from the TCGA (the cancer genome atlas) project \cite{14}. We scope our work to the renal cell carcinoma (RCC) which is one of the most common malignant tumors in adult kidney type, which has three common histologic subtypes in the dataset: clear cell (TCGA-KIRC), papillary (TCGA-KIRP), and chromophobe (TCGA-KICH). In order to validate the robustness of our framework, we apply the same technique to the independent renal cell carcinoma dataset from the local hospital as well. Each case in the dataset has its corresponding pathological report and clinical information. The reports are made after pathologists observed the histopathological sections stained with hematoxylin and eosin (H\&E) \cite{15} under a microscope to make the diagnoses by inspecting the morphological characteristics of histology slides. With the pathology report, clinicians—with overall understanding of patients’ symptoms—can make informed and precise decisions about treatments. We can verify the prediction results of the model with the corresponding content of the pathology report to test the accuracy of our framework \cite{wu2020structured}.

In this work, we first constructed large, high-quality, fine-grained annotated renal cancer datasets to address the lack of large, accurately annotated datasets in the study of digital pathology research. Based on this dataset, a deep learning model is used to detect cancer regions, classify tumor subtype and grade on the whole-slide image, and finally generate a precise whole-case report. We test our framework on the dataset from the hospital and invite experienced pathologists to help additional validation. Our work could fundamentally alleviate the problem of long training cycles for pathologists, and further research will help demonstrate that the same approach for other types of cancer (such as prostate cancer, lung cancer).

The organization of this paper is as follows: the second part discusses the dataset of pathological images and describes our experimental process in detail. In the third part, the results section introduces the performance of our framework, and in the conclusion part, we discuss the experimental results and the future work.

\section{Materials and Methods}
In order to detect cancer regions, classify cancer subtypes and cancer grades on the whole-slide images, and finally generate a whole-case report, a group of population with similar histology in different dataset project is needed for each case.

\subsection{Data sources and annotations}
We selected renal cell carcinoma in our study and evaluated our frameworks on two datasets. It is worth mentioning that the whole-slide images we used here are the digital diagnostic slides of each patient, each patient was given a corresponding pathological image. Our dataset comes from NCI Genomic Data Commons \cite{40}, which provides an online research platform for uploading, searching, viewing and downloading cancer-related data. All free whole-slide images of kidney cancer are uploaded from this source. The first dataset derived from the TCGA database we name as TCGA RCC. This dataset we used contains three TCGA projects (i.e., KIRC, KIRP, KICH) and totally has 667 WSIs. All of the whole-slide images were scanned at 40x magnification, and the data were cleaned by experienced pathologists before they were annotated to remove the blurred, non-cancerous images.

As part of this work, we introduced a new RCC dataset which is obtained from the Department of Pathology, the First Affiliated Hospital of Xi’an Jiaotong University, China. We name as LH RCC. We totally collected 632 WSIs from 153 patients in hospital, which is about 50 patients for each subtype. These WSIs are paraffin-embedded tissue sections and scanned at both 20x and 40x magnification by the digital slide scanner of KFBIO. The labels provided by the TCGA and Local hospital datasets were used as our gold standard. Those labels were the result of a consensus as explained by the data curator: first, the submitting institutions were asked to review each sample prior sending it to confirm the diagnosis. Then, a slide from the sample was reviewed by an expert pathologist. In the event of a disagreement, the slide would be reviewed by one or more other expert pathologists.

\begin{figure*}[]
\centering
\includegraphics[width=120mm]{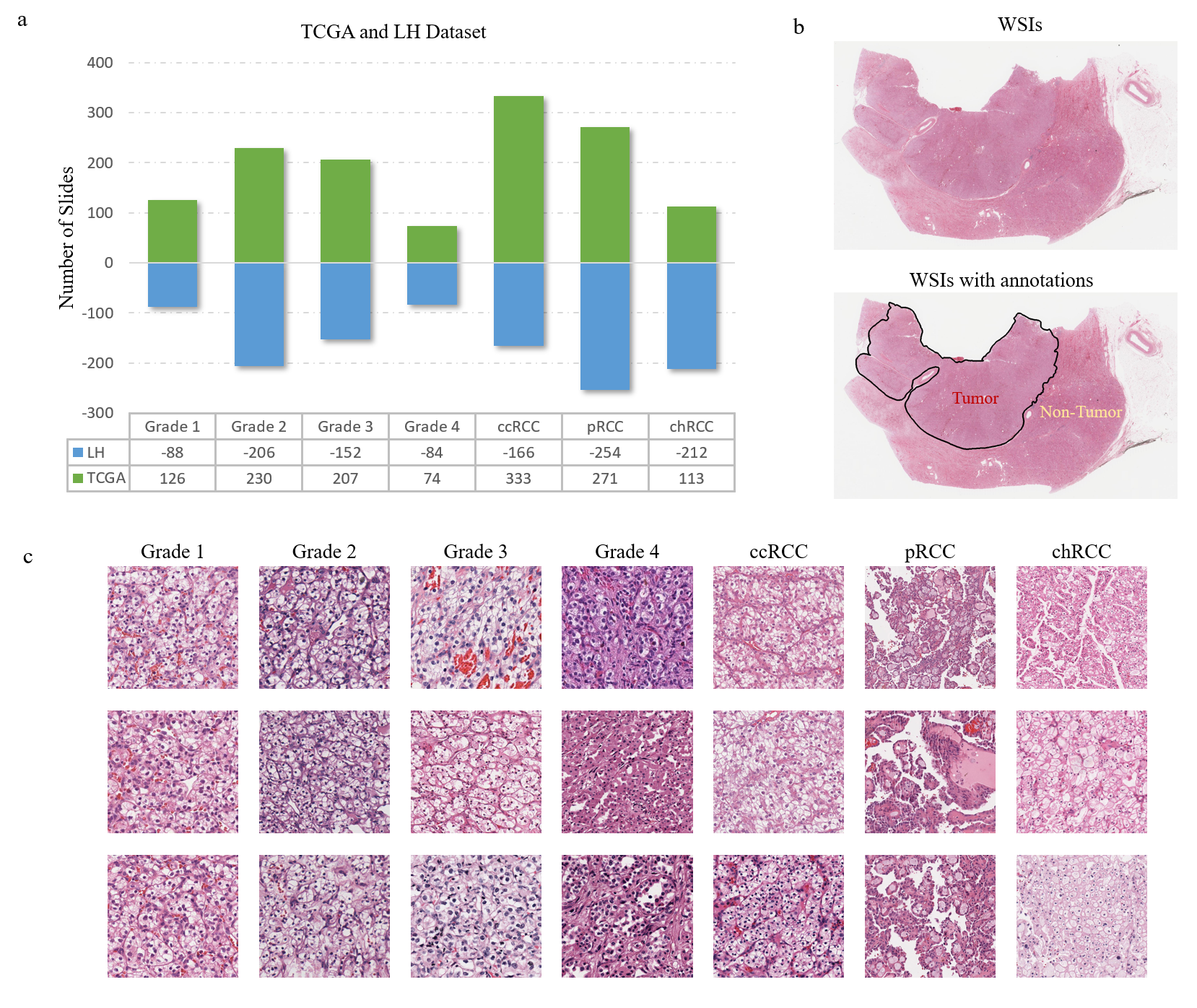}
\caption{Introduction of the two datasets. (a) The bar chart shows and counts the detailed of different grades and different subtypes in the two datasets. (b) The original digital pathological whole-slide image, a whole-slide image annotated by the pathologist, distinguishes the cancer area from the normal area, and provides fine-grained annotations of subtypes and grades. (c) Examples of patches in different types include three different subtypes and four different grades.}
\label{f1}
\end{figure*}

\begin{figure*}[]
\centering
\includegraphics[width=180mm]{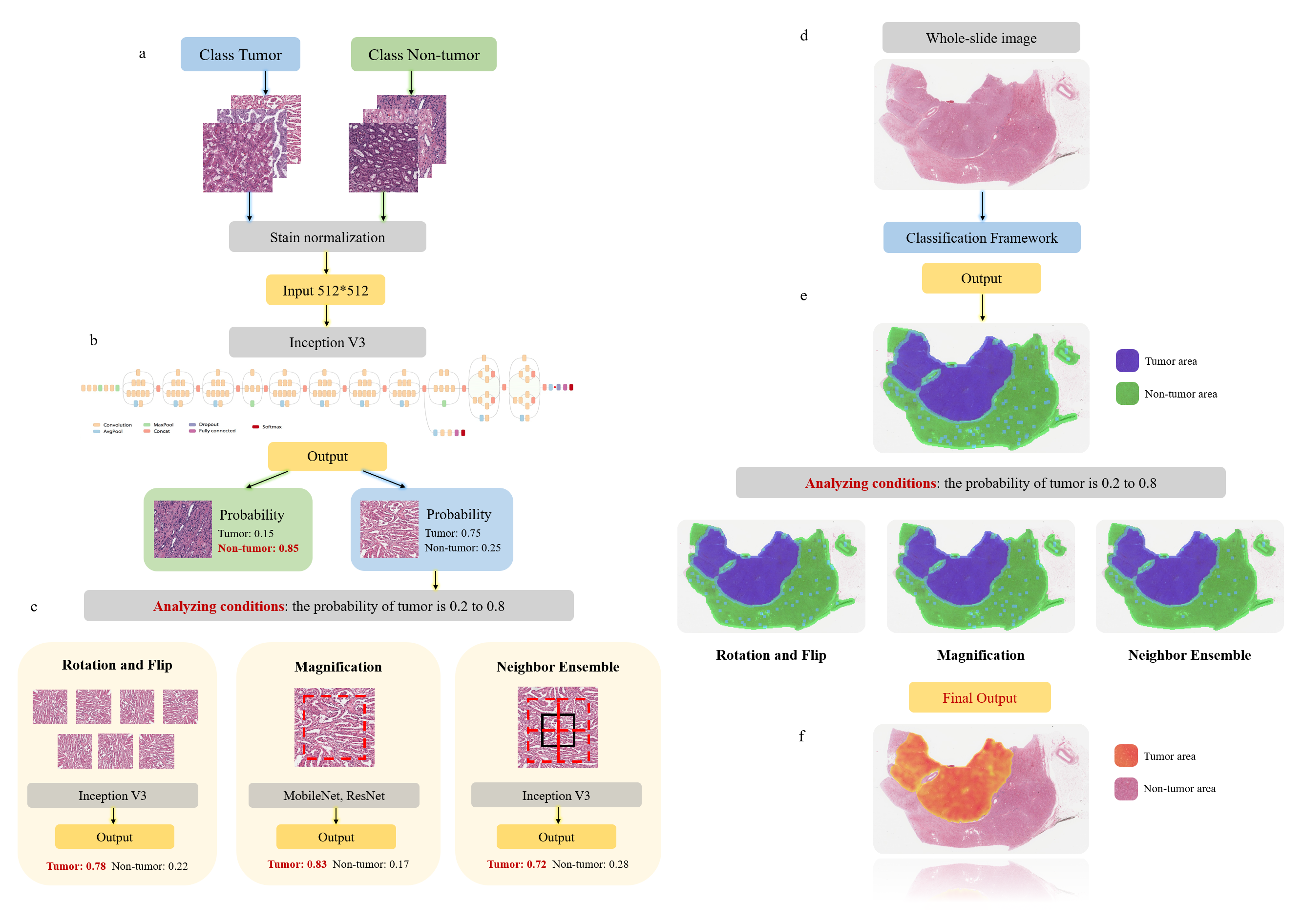}
\caption{The convolution network structure of the tumor region detection subtask in the framework and a visualization example of each step. (a) The input of the network is different types of images (cancer patch and normal patch) for training, 512*512px in size, with stain normalization. (b) We used the Inception V3 from Google with initialized our network parameters to the best parameter set that was achieved on ImageNet competition. The output expresses the probability of different classes (Tumor and normal), If the probability of patch currently predicted as cancer area is less than 0.2 or greater than 0.8, the results will be output. (c) If the probability of patch prediction as cancer area is between 0.2 and 0.8, three different methods (Rotation and flip, magnification, neighbor ensemble) are used for second judgment, and the final output is obtained by voting on the three results. (d) The selected WSI staining was not ideal, with obvious cutting edges and artifacts. (e) After visualizing output, it could be seen that there were obvious "wrong judgment" areas except the cancer areas. We used three methods to make the second prediction and vote to integrate the results. (f) The final tumor area map is generated by overlaying the probability map on the WSI slide to facilitate the pathologist's interpretation of the detected tumor area. }
\label{f2}
\end{figure*}

\subsection{Annotation principles}
OpenHI \cite{41} is an open-source annotation and processing software which specially designed for histopathology images. Now, it supports several types of annotation and can record the time of annotation. We performed all operations including tumor region selection and region annotation on the OpenHI platform. For the tumor region detection and subtype recognition subtasks of this framework, two experienced pathologists first selected the tumor area for the whole-slide images, and we could take patches in the normal and tumor areas respectively. After selecting the tumor regions, pathologists annotated 123, 88, and 46 WSIs for ccRCC, pRCC, and chRCC respectively in the TCGA dataset, and four well-trained annotators were invited to make the complete region annotation on the remaining WSIs for the tumor region and three subtypes. For the dataset in the local hospital, we randomly selected ten patients of each subtype, 180 WSIs in total, then asked two pathologists to do the same annotation on these WSIs. It is worth mentioning that all the cases in these two datasets have corresponding pathology reports and clinical records which contain the diagnostic subtype labels and the grading score. All the annotation we have done is to annotate the cancerous and non-cancerous regions.

For the grades classification subtask in this framework, from the WHO guideline we can know that, the renal cell in tumor area of the whole-slide images are divided into 4 grades based on the ISUP grading standard \cite{42}. This indicator is particularly important in the diagnosis of clear cell carcinoma. We chose 200 slides image of renal cell carcinoma and chose 10 diagnostic areas at 40X magnification on each slide in both TCGARCC dataset and part of the LHRCC dataset at 40x magnification. The size of each selected region is 512*512 pixels. Six rigorously trained annotators which were divided into two groups to annotate the chosen regions. After completing the data annotation, we compute the inter-rater reliability between different annotators in a group. According to the interpretability of kappa statics \cite{43}, the confidence of different annotators is at moderate level. In order to achieve high quality annotation, the class of each instance was decided based on the decision of the majority which consist of three annotators in one group. The details of these two datasets are shown in Figure 1.

\subsection{Stain normalization}
The normalization of haematoxylin and eosin stain was made at the training and validation stages using a reference image and the Macenko algorithm \cite{44}.

\subsection{The proposed framework}
Figure 2 depicts our data analysis workflow for tumor region classification subtask, Figure 3 depicts our subtask for tumor subtype and grade classification. With the following sections describing the information for each major step in our framework.

\subsection{Tumor and Non-tumor region classification}
For the tumor area detection task, we used part of the annotated data in the two datasets as the training data, and use the rest of the data as the validation data. In each annotated WSI, 5 patches were selected for tumor region and non-tumor region respectively as two different classes. After staining normalization, the patches were input into the convolutional network model. In the selection of the model, we trained several different image classification neural networks (VGGs, ResNet, MobileNet and InceptionV3). According to the result we find that Inception V3 provided the best performance and was used for all the final experiments.

We used 80\% of those patches for training, 20\% for validation and testing. We based our model on Inception V3. This architecture makes use of inception modules. The initial 5 convolution nodes are combined with 2 max pooling operations and followed by 11 stacks of inception modules. The architecture ends with a fully-connected layer and then a SoftMax output layer. The last three layers of the model are discarded, and then the result of the bottleneck layer is used as the feature extraction result of the model. After obtaining the feature vector, it needs to be input into a fully-connected layer for classification. We initialized our network parameters to the best parameter set that was achieved on ImageNet competition. We use exponential decay learning rate:

$global_step$ is a counter, counting from 0 to the number of iterations of training; $learning_rate$ is the initial learning rate; $decayed_learning_rate$ decays as $global_step$ increases; $decay_steps$ is used to control the decay speed.

Inspired by the work of prostate cancer classification [8], the accuracy of convolutional network can be improved by 0.7\% during the training and validating of stain normalized patches. The result of the convolutional neural network calculation is the probabilistic quantitative information of different categories of patches. According to the pathologist's experience in diagnosis, the probability of tumor diagnosis in the patch with malignant cancer is more than 0.9. In the diagnosis results of our convolutional neural network, three different augmentation strategies were added to our framework to improve the accuracy of the model during validation and reduce the misdiagnosis that should not occur, in order to avoid the wrong diagnosis of normal tissue as a tumor region due to incorrect threshold selection. The condition to trigger the data enhancement strategy is that the probability of the current patch being diagnosed as tumor tissue by the convolutional neural network is between 0.2 and 0.8.

The first method of data enhancement is to create seven derivations (three spins and four reversals) on the patch (Rotation and flip), test the derived patch with the convolutive network, calculate the tumor probability of the eight patches including the original patch, take the median and confirm the final classification of the tumor probability with 0.5 as the threshold. The second method uses the convolutional network model with the same principle of high accuracy training to select a higher amplification factor for the current patch training (Magnification), which predominately through additional yield on tumor tissue recognition accuracy. The third method to improve diagnosis accuracy is to find the auxiliary environment patch which surrounds the current patch (Neighbor ensemble). For the current patch, we choose its geometric center and get four environment patches at the same magnification and size in the region near the geometric center, then using the convolution network training four environmental patches to calculate the probability of diagnosis of tumor. We average the four different results as the final probability and take 0.5 for the threshold. The reason for adopting this approach instead of selecting eight patches around the current patch is to maximum retain the semantic information in the current image and improve the computing speed as much as possible.

In our experiment, training with the above three methods of data enhancement for each patch does not improve the final diagnostic accuracy, it is worth mentioning that we do not enhance the data for each patch in the model but only carry out secondary verification for the patches that meet the requirements (probability between 0.2 and 0.8).

\subsection{Tumor subtype and grading classification}
For both subtype and grade classification subtasks, we confirmed the selected tumor areas in the previous section to determine the subtypes and grades of the cancer areas. We meshed the tumor areas and ensured that each area validated experimentally included tumor areas and tumor boundaries. The classification network we used is similar with the network for tumor region classification. The difference is that, for subtype classification task, our input and output correspondingly become the probability of images and patches of three different subtypes belonging to the three categories respectively. For the grade classification task, our input and output are the images corresponding to ISUP grade 1 to grade 4 and the probability values obtained.

In our experiments for subtype classification, we have well-delineating tumor architecture and trained models for classification using three classes (ccRCC, pRCC and chRCC). For the grading classification we trained models using four classes (G1, G2, G3, G4). According to the experience and the observation during the annotation process of the pathologists, there are almost one types of subtypes in one patch, but different grades of tumor cells will appear in one patch. As a training dataset, we only used patches with pure and clear patterns from tumor areas, each patch contained only one subtype and one grade, so as to reduce the differences among observers. Three classes for pure subtype (ccRCC, pRCC and chRCC) and four classes for pure grade (G1, G2, G3, G4) were used for training. We used argmax strategy for the patches in the final classification of subtypes, however we found that the best result for ISUP grading do not stem from the final classification of patches using the argmax strategy. Patches with different grade of tumor cells represents the intermediate representation of tumor structure, not just the mixture of different grades. Such statistics can well describe the tumor differentiation of renal cell carcinoma and more accurately summarize the overall ISUP grade of the cancer area from the grading probability of individual patches. During validation, the grading experiments included the calculation of the percentage of single grade. In the validation dataset, the International Society of Uropathology (ISUP) grade groups 4 were pooled.

\begin{figure*}[]
\centering
\includegraphics[width=180mm]{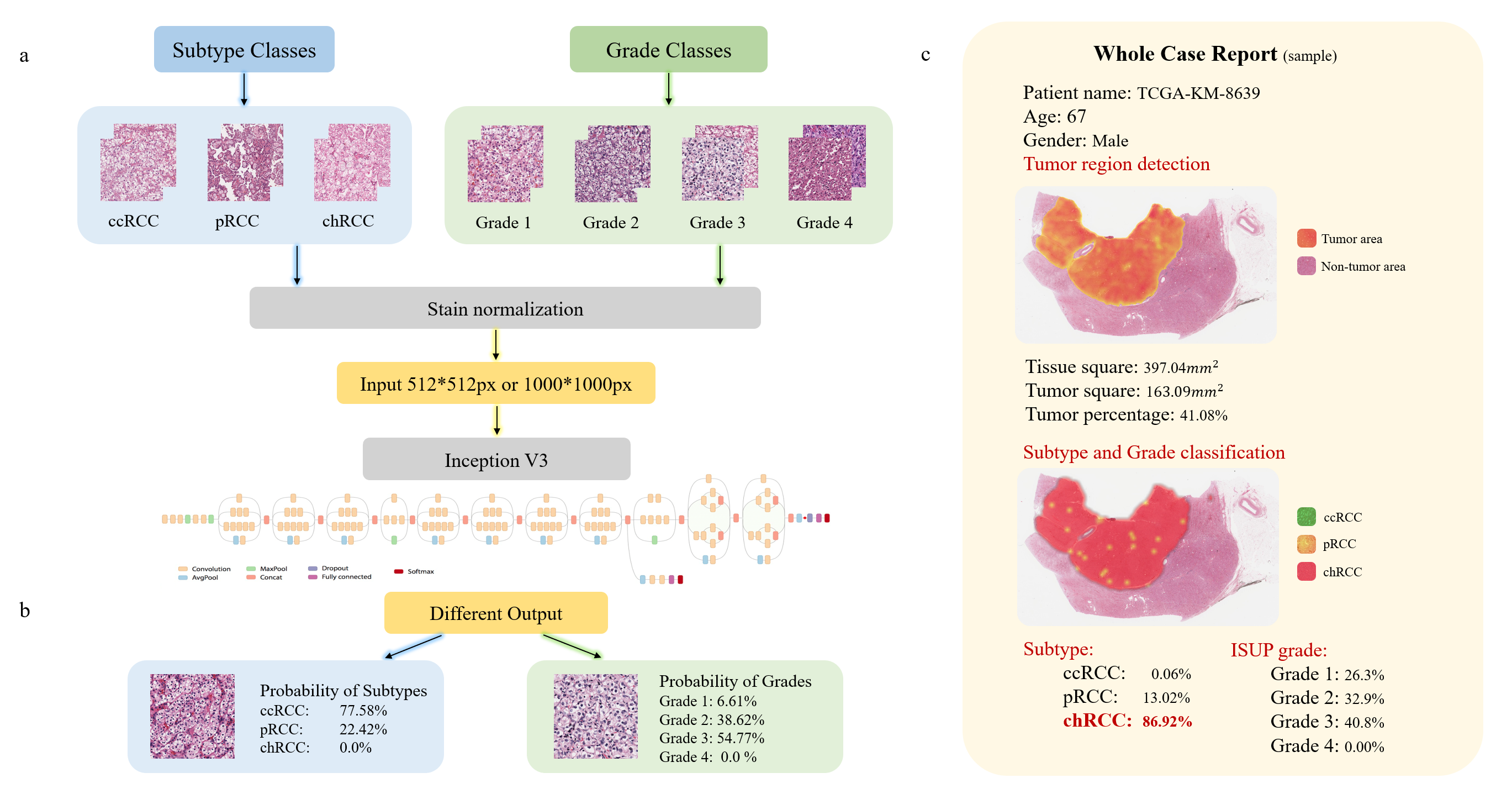}
\caption{The convolution network structure of the subtype and grading subtask in the framework. (a) Similar to the classification of cancer regions, we used similar network structures for subtype classification and classification. Subtype classification included three different subtypes, and classification included four different ISUP grades in clear cell renal carcinoma (ccRCC). (b) For subtype classification, the result is the probability that each patch belongs to different subtypes. For grade classification, first judge whether it is grade dour, and then calculate the probability that each patch belongs to three grades. (c) The Whole Case Report contains the detailed information of patients in the current Whole-slide image by our framework: (1) The tumor detection heatmap; important WSI metrics (tissue Square, tumor Square and tumor) are generated from tissue classification results; (2) The subtype classification heatmap; the statistical probability of different subtypes in the Whole-slide image; (3) Statistical information of ISUP classification.}
\label{f3}
\end{figure*}

\subsection{Whole-case report generation}
From the result of the classification convolutional network model, we can get the tissue area of the whole-slide image, the area of tumor area, the proportion of tumor area, the detailed information of subtypes, and the detailed information of classification. Pathological report can clarify the diagnosis of the disease and provide accurate basis for doctors to treat the disease, evaluate the prognosis and explain the symptoms. Therefore, the department of pathology should not only ensure the timeliness of the report, but also ensure its accuracy. Precise calculation of the different metrics which is naturally difficult for human pathologists could be used for generation of the whole case report.

\subsection{Data availability}
All the relevant data used for training during the current study are available through the Genomic Data Commons portal (https://gdc-portal.nci.nih.gov). These datasets were generated by TCGA Research Network (http://cancergenome.nih.gov/), and they have made them publicly available. Other datasets analyzed during the current study are available from the corresponding author on reasonable request.

\section{Result}

\subsection{Classifying tumor versus non-tumor tissue}
Among the perfect accuracy convolutional network in image classification, the best convolutional neural network architecture which perform the best accuracy was InceptionV3. The size of each patch representing different magnifications were tested from range ×20 to ×40. The best result for classification were achieved for the magnification of 500 × 500 px (patch size of 128 × 128 px scaled from originally generated 600 × 600 px patches). The classification accuracy of the native convolutional network was 93.8\% for training and 92.2\% and 92.7\% for validation dataset TCGARCC and LHRCC respectively, and could be further improved through the additional deep learning-based strategies.

For each patch, the convolutional neural network calculates the quantitative information of probability belonging to a certain class (tumor or non-tumor).In most of the tumor patches, the probability of tumor determination is 0.90-1.0.However, if the patch contains normal tissue, if the tumor lesion is small or well differentiated, or if there is mis operation and artifact in the process of making slides by the pathologist, it will be less likely to judge the patch as a tumor. In the "Methods section", we proposed three strategies to enhance the output of convolutional neural network to further improve the overall accuracy of the framework. The trigger factor for the secondary diagnosis and analysis and calculation of each patch is that the probability of belonging to the tumor category output by convoluted network is between 0.2 and 0.8. From the experiment we find that 3.9\% patches in the verification dataset TCGA and 3.6\% of validation dataset Local hospital need to have the secondary validation by the three different strategies.

Among our three strategies, the first strategy which creating seven derivations (three spins and four reversals) on the patches (Rotation and flip) shows impressive result and greatly improves the accuracy of the model by 0.73\%. The second strategy which selecting a higher amplification factor for the current patch training (Magnification) also improves the accuracy of the model by 0.68\%. The third strategy which finding the auxiliary environment patch which surrounds the current patch (Neighbor ensemble), by analyzing the four patches around the model, greatly reduces the time and cost required for calculation, and also achieves good results in tumor image classification. As shown in Figure 2, a complete workflow was developed for WSI analysis using trained convolutional network to realize model prediction and enhancement strategy for part of patches and provide important tumor indicators for pathological reports. We selected a pathological image with unsatisfied staining, obvious artificial cut marks and irregular artifacts for the framework process display, and the final tumor thermogram was superimposed on the WSI slide to facilitate the pathologist to observe and interpret the detected tumor area. We show that the output accuracy of the model can be significantly improved through P8, ENS and ENV strategies. In tumor area detection, important WSI indicators (the area of WSI, the area of the tumor area, and the proportion of the tumor area) are generated from the classification results and used to generate complete case reports that cannot be accurately calculated by the pathologist during the diagnosis.

\begin{table}[]
\begin{center}
\caption{Performance of different segmentation models for patch-level on the data set}
\begin{tabular}{|l|l|l|l|}
\hline
\multicolumn{1}{|c|}{Deep learning model} & \multicolumn{3}{c|}{Cancer Region Detection} \\ \cline{2-4} 
\multicolumn{1}{|c|}{}             & ccRCC         & pRCC         & chRCC         \\ \hline
ResNet-34 \cite{resnet}                 & 0.87          & 0.82         & 0.84          \\ \hline
DenseNet \cite{densenet}                & 0.89          & 0.83         & 0.89          \\ \hline
U-Net \cite{unet}                       & 0.81          & 0.78         & 0.84          \\ \hline
DeepLab v2 \cite{chen2017deeplab}       & 0.91          & 0.84         & 0.87          \\ \hline
Inception v3 \cite{inception}           & 0.93          & 0.86         & 0.91          \\ \hline
\end{tabular}
\end{center}
\end{table}

\subsection{Tumor subtype classification}
Similar to the classification of tumor tissue regions, we adopted the same convolutional network structure in the classification of subtypes and got the probability of three classes (ccRCC, pRCC and chRCC). Figure 3 shows the convolutional network and data output. The patch size we selected was 1000 × 1000 px. In the process of subtyping tumors, the selection of patch size needs to consider the model separately, because the characteristics of subtypes need to be considered in the field of vision. If the patch is too small, then the characteristics of papillary renal cell carcinoma cannot be well presented. We trained on two datasets and selected cancer images that were not used for training for verification. Since each whole-slide image has the corresponding basic information of the patient which provides labels on subtypes and grades, the results our model predicted can be compared with ground truth. In Figure 3, we use the same pathological images as in Figure 2, and it can be seen that only the cancerous region is separated by different colors, and different colors represent the probability of a single patch being output through the convolutional network. Precise calculations of different indicators (the proportion and area of each subtype) are naturally difficult for human pathologists and can be used to generate whole-case reports.

\begin{table*}[]
\begin{center}
\caption{Performance of different classification models for patch-level on the data set}
\begin{tabular}{|l|l|l|l|l|}
\hline
\multicolumn{1}{|c|}{Deep learning model} & \multicolumn{4}{c|}{Classification}                                       \\ \cline{2-5} 
\multicolumn{1}{|c|}{}                                     & 3-Class Subtyping & 4-Class Subtyping & 3-Class Grading & 4-Class Grading \\ \hline
ResNet-34 \cite{resnet}                 & 0.78              & 0.82              & 0.84            & 0.82            \\ \hline
DenseNet \cite{densenet}              & 0.86              & 0.89              & 0.87            & 0.84            \\ \hline
U-Net \cite{unet}                       & 0.77              & 0.78              & 0.81            & 0.79            \\ \hline
DeepLab v2 \cite{chen2017deeplab}       & 0.86              & 0.89              & 0.86            & 0.87            \\ \hline
Inception v3 \cite{inception}            & 0.87              & 0.90              & 0.91            & 0.88            \\ \hline
\end{tabular}

\end{center}
\end{table*}

\subsection{Tumor grading classification}
Tumor classification and subtype classification are multi-classification problems, so we chose the same convolutional network structure to obtain the probability of four types belonging to four categories (G1, G2, G3 and G4). In Figure 3, we showed the workflow of the two subtasks together. Different from the subtype classification task, based on the understanding that the pathologist indicates the cancer level is continuous. In the classification of tumor grades, to maximize the retention of the phenotypes of the different tumor cells, especially the hierarchical tumor cells, are described in the WHO guideline, tumors showing extreme nuclear pleomorphism and/or containing tumor giant cells and/or the presence of any proportion of tumor showing sarcomatoid and/or Rhabdoid differentiation. In the preprocess of treatment, the tumor area was cut into patches of 1000px*1000px, and then the probabilities were analyzed and calculated respectively through the convolutional network. When grading and classifying patches in each cancer region, we first adopted dichotomy and divided each patch into grade IV cancer cell nuclei and non-grade IV cancer cells. For non-grade IV cancer patches, the three-classification model is adopted, and the results are the probability value of each patch at grade I II III respectively. For the result of classification task, each patch needs to keep its own probability value which belong to different classification categories, and in the calculation of the cancer grade of the whole-slide image, the whole tumor area should be aggregated for calculation, instead of each patch being nominated to an independent classification (each patch in the subtype classification has its own independent corresponding subtype). By aggregating the entire tumor area for calculation, information about cancer cells of different grades can be retained to the greatest extent. 

\subsection{Whole-case report generation}
A detailed pathological report can help clinicians to carry out more targeted treatment measures, which is of great significance in both pathology and clinical medicine. The size and proportion of tumor areas in the whole tissue image, the size and proportion of different subtypes in the whole image and the proportion of cancerous areas are accurately calculated by convolutional network classification, and the cancer level area and proportion of tumor areas are calculated by statistical method. It is difficult for human pathologists to quantify these different indicators, which can be used to generate whole-case reports.

\section{Discussion}
In artificial intelligence for cancer diagnosis, deep learning is applied to histological image detection of tumor tissues and standard subtype classification, grading, and other pathological tasks in the whole-slide images. The main direction of research is semantic-based image segmentation using different methods for patch-level classification of histological structure. However, the lack of publicly annotated datasets slows the development of artificial intelligence in pathology. We have created a large training dataset from nearly 1300 slides representing different patients in different institutions and public datasets with extensive, high-quality annotations. We agreed that building datasets with accurate annotations are necessary to improve model accuracy using deep learning models, and we will continue to expand our data sets in future work to include more cancer types and cancer images.

This study is a new application of deep learning algorithms in renal cell carcinoma to recognize and classify tumor tissues in histological digital slides. Our study shows that a convolutional neural network can be used for histopathological slides in the diagnosis of renal cell carcinoma, which correctly distinguished normal tissue from tumor tissue, identified subtypes and grades of renal cancer with high accuracy, and achieved a sensitivity and specificity comparable to that of the pathologists.

In analyzing the prediction results, we noticed that several whole-slides images were misclassified by both the neural network and pathologists. We consulted pathologists about the occurrence of this condition, after several diagnoses of experienced pathologists, some of the erroneous results were due to errors in the TCGA diagnosis, as well as different results due to different grading methods. Similarly, we selected some of the pathologist's misclassified images in the diagnosis result, then we used our framework to predict the diagnosis, and showed that our diagnosis was identical to the ground truth, which shows that our framework could help auxiliary diagnosis to the pathologists and make the correct decisions.

In our work, we used three different strategies to improve the accuracy of tumor detection. Tumor area detection is critical in our framework because the subsequent subtypes and grading classification work is predicated on the tumor area These strategies need to add calculation amount for the patch that meets the conditions, but they improve classification accuracy, which is an acceptable increase in calculation amount. Besides, the post-training of deep learning-based models with new data can greatly improve the accuracy of models, which is very important for the continuous development of models.

There are limitations to our work. First of all, we asked the pathologists to perform multiple rounds of high-quality renal carcinoma pathologic imaging for one year, but the quantity is far from enough. We need to incorporate new data sets (more images of kidney cancer and other types of cancer) into our work to further validate and develop the integrity of our work. Second, further research should include work on the classification of the types in PRCC and the classification of the fourth-grade phenotypes in CCRCC (rhabdomyosoid, sarcomatoid, necrotic, and hemorrhagic). In papillary renal cell carcinoma, different types (type I and type II) have different morphological characteristics, which contain different prognostic significance. For example, PRCC type II shows a higher grade of pathologic T staging and ISUP grade, and this type is more prone to sinus/perirenal adipose infiltration and sarcomatoid differentiation. There is a lot of morphological overlap between type I and type II in PRCC, and type II patients have a worse prognosis. It is of great significance for the prognostic evaluation of PRCC to correctly carry out tissue-credit type.

Overall, our work shows that convolutional neural networks can help pathologists classify whole-slide images of renal cell carcinoma and generate whole-case reports in deep learning. Pathologists can use this information to treat patients more specifically. We hope to apply this framework to identify papillary renal cell carcinoma and other kind of cancer in future work. Also, identifying some critical features of grade IV clear cell carcinoma (rhabdomyosoid, sarcomatoid, necrotic, and hemorrhagic) and some critical diagnostic features summarized by pathologists is also important to the diagnosis and prognosis of the pathologic diagnostic procedure. In addition to renal cell carcinoma, our model can be applied to other cancers in the future. By expanding the robustness of our model to recognize a wider range of histological features through various types of cancer, complete and accurate pathologically assisted diagnosis can be made, effectively helping pathologists in daily tasks and specific cases.

\section{Conclusion}
In this study, we developed a precise framework for analyzing whole-slide images of renal cell carcinoma patients based on a deep learning model, which has the potential to achieve pathologist-level accuracy and interpretable auxiliary diagnosis. This framework is realized in the pipeline of digital pathology which provides all the relevant tumor metrics and all the necessary information for generating a pathology report and assists the pathologist in their daily work. Further validation of diagnostic histopathology workflows in the real world is necessary. The pathologist can understand these prediction result from the deep learning method when performing the second examination and visual examination. Our approach is data-agnostic. With the success of deep learning, we believe that our approach has a strong universality for learning the complex tissue structure of different types of cancer. Further research will help to demonstrate its efficacy against other types of cancer, such as lung cancer, breast cancer and prostate cancer. Exploring the use of multiple types of information for diagnosis is clinically necessary.


\section*{Acknowledgment}
This work has been supported by National Natural Science Foundation of China (61772409); the Innovative Research Group of the National Natural Science Foundation of China (61721002); and the consulting research project of the Chinese Academy of Engineering (The Online and Offline Mixed Educational Service System for “The Belt and Road” Training in MOOC China);  The results shown here are in whole or part based upon data generated by the TCGA Research Network: https://www.cancer.gov/tcga.



%\bibliographystyle{IEEEtran}


% \end{thebibliography}
% \bibliographystyle{IEEEtran}
% \bibliography{reference}


\end{document}